\documentstyle[aps,prb,epsf,multicol]{revtex}
\begin{document}

\newcommand{\oldc}{{\hat c}}
\newcommand{\newc}{{\hat d}}
\newcommand{\oldS}{S}
\newcommand{\newS}{T}

\title{Spin polarons in triangular antiferromagnets}
\author{Matthias Vojta}
\address{
  Institut f\"{u}r Theoretische Physik,
  Technische Universit\"{a}t Dresden, D-01062 Dresden, Germany \\
}
\maketitle

\begin{abstract}
The motion of a single hole in a 2D triangular antiferromagnet is investigated 
using the $t$-$J$ model. 
The one-hole states are described by strings of spin deviations around the hole.
Using projection technique the one-hole spectral function is calculated.
For large $J/t$ we find low-lying quasiparticle-like bands which 
are well separated from an incoherent background by a gap of order $J$.
However, for small $J/t$ this gap vanishes and the spectrum becomes broad over
an energy range of several $t$.
The results are compared with SCBA calculations and numerical data.
\end{abstract}

\widetext
\begin{multicols}{2}
\narrowtext



Since the discovery of high-temperature superconductivity, charge carriers
in doped antiferromagnets (AF) have been studied intensively. 
A reliable description of the hole motion is important for the understanding of
the low-energy charge dynamics in the copper-oxide planes of the cuprate
superconductors.
A large number of numerical and analytical studies indicate that
a single hole in an AF spin background has nontrivial
properties:
The spectral function consists of a pronounced coherent peak at the bottom of the 
spectrum and a incoherent background at larger energies.
The coherent peak can be associated with the motion of a dressed hole,
i.e., a hole surrounded by spin defects ("spin polaron")
~\cite{Dagotto94,Strings,Trugman88,Riera97,Vojta98,MaHo}.

Materials with spin arrangements on 2D non-square lattices have also been 
synthetized.
For example, experiments suggest the realization of a
triangular spin-$1 \over 2$ AF in NaTiO$_2$~\cite{Takane94},
as well as in surface structures~\cite{Surfaces} such as 
K/Si(111):B. Delafossite cuprates RCuO$_{2 + \delta}$,
with R a rare-earth element, have Cu ions sitting on a triangular
lattice~\cite{defa}.
Furthermore, recent results in the context of organic
superconductors indicate that $\kappa$-(BEDT-TTF)$_2$X, where X is an ion,
may be described by a half-filled Hubbard model on an anisotropic triangular 
lattice~\cite{BEDTTTF}.

Although most of these materials do not contain a 
finite concentration of holes or electrons away from half-filling, 
it is of fundamental theoretical interest to study the hole dynamics 
in this environment.
In this paper we adress the motion of a single hole in an otherwise 
half-filled system. 
The one-hole spectral function for this case corresponds directly to the
result of an angle-resolved photoemission experiment (ARPES) on the undoped 
compound.

\begin{figure}
\epsfxsize=7.5 truecm
\centerline{\epsffile{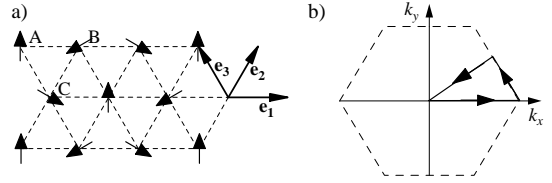}}
\caption{
a) The classical ground state of the triangular AF 
   together with the three unit vectors ${\bf e}_i$ of the lattice.
   The 120$^o$ spin order defines three sublattices A, B, and C.
b) Brillouin zone for the triangular system with
   the path in momentum space for the following figures.
}
\label{FIG_LATTICE}
\end{figure}

We assume that a triangular AF doped with holes is well described 
by the $t$-$J$ model on a triangular lattice,
\begin{equation}
H\, =\,
    - t \sum_{\langle ij\rangle \sigma}
      (\oldc^\dagger_{i\sigma} \oldc_{j\sigma} +
       \oldc^\dagger_{j\sigma} \oldc_{i\sigma})
    + J \sum_{\langle ij\rangle} \ {\bf\oldS}_i \cdot {\bf\oldS}_j
\label{H_TJ}
\end{equation}
in standard notation. 
Note that, in contrast to the square lattice, there is no electron-hole
symmetry for the triangular lattice $t$-$J$ model, i.e., the physics
depends on the sign of $t$.
For half-filling this model reduces to the triangular Heisenberg
antiferromagnet (THAF) with spins $1 \over 2$\cite{Fazekas74}.
It is by now widely believed
that its ground state possesses magnetic long-range order
\cite{Singh92,Bernu94} which can be described as the 120$^o$ order 
being the ground state of the classical spin model (Fig. \ref{FIG_LATTICE})
with additional quantum fluctuations.

To investigate the hole motion we consider a one-particle Green's function 
describing the creation of a single hole with momentum $\bf k$
at zero temperature:
\begin{equation}
G({\bf k},\omega) \;=\; \sum_\sigma
  \langle\psi_0^N| {\hat c}_{{\bf k}\sigma}^\dagger {1 \over {z-L}}
               {\hat c}_{{\bf k}\sigma}
  |\psi_0^N\rangle
\label{HOLE_GF}
\end{equation}
where $z$ is the complex frequency variable, $z=\omega+i\eta$, 
$\eta\rightarrow 0$. 
The quantity $L$ denotes the Liouville operator defined by $LA = [H,A]_-$ for arbitrary
operators $A$.
$|\psi_0^N\rangle$ is the ground state of undoped system, i.e., with $N$ electrons on $N$
lattice sites.

The one-hole problem in the triangular lattice
has up to now analytically only been studied 
\cite{Azzouz,Apel} using self-consistent Born approximation (SCBA).
The SCBA (without vertex corrections) neglects spiral
loops in the hole motion (Trugman processes\cite{Trugman88})
since they are formally crossing diagrams.
However, for the triangular lattice
these processes are expected to be more important than for the square
lattice since only three hopping steps are necessary for one loop.
Here we prefer another analytical approach which is based on the 
picture of the spin-bag quasiparticle (QP) or magnetic polaron 
\cite{Strings,Trugman88,Riera97,Vojta98,MaHo}.
The spin deviations surrounding the hole are described by a set of path operators
\cite{Strings,Trugman88,Riera97,Vojta98}.
The correlation function (\ref{HOLE_GF}) can be evaluated using 
a cumulant version \cite{BeckFul88,BeckBre90} 
of Mori-Zwanzig projection technique\cite{Projtech}.



In the calculations of this paper the ground state $|\psi_0^N\rangle$ of the undoped
THAF is assumed to be long-range ordered \cite{Singh92,Bernu94}. 
Its description is based on an expansion around the classical ground state 
$|\phi_{\rm cl}\rangle$ of the triangular antiferromagnet, see Fig. 1.
Translational invariance and rotation invariance in spin space are
spontaneously broken; we choose the coordinates that the spins in $|\phi_{\rm cl}\rangle$ are
arranged in the $x$-$z$ plane.

For the analytical treatment we locally rotate all spins 
to transform the classical ground state into a formally ferromagnetic 
state. The rotation is carried out
around the $y$-axis by an angle of ${\bf Q} \cdot {\bf R}_i$ with
${\bf Q} = ({8 \over 3} \pi,0)$ at site $i$. 
The "original" electron operators $\oldc^{(\dagger)}$ and the "new" operators
$\newc^{(\dagger)}$ in the rotated basis are related through:
\begin{equation}
\oldc_{i\sigma} \;=\; 
\cos \frac{{\bf Q} \cdot {\bf R}_i}{2} \: \newc_{i\sigma} +
\sin \frac{{\bf Q} \cdot {\bf R}_i}{2} \: \newc_{i,-\sigma}
 \, .
\label{OLDNEWTRAFO}
\end{equation}
The resulting Hamiltonian $H = H_0 + H_1$ has the following form:
\begin{eqnarray}
H_0 &\;=\;&
    + {t \over 2} \sum_{\langle ij\rangle \sigma}
      (\newc^\dagger_{i\sigma} \newc_{j\sigma} +
       \newc^\dagger_{j\sigma} \newc_{i\sigma})
    \,+\, {J \over 2} \sum_{\langle ij\rangle} \ \newS_i^z \newS_j^z
 \nonumber \\
H_1 &\;=\;&
    - \, {t \sqrt{3} \over 2} \sum_{\langle i\rightarrow j\rangle \sigma} \sigma
      (\newc^\dagger_{i\sigma} \newc_{j,-\sigma} -
       \newc^\dagger_{j\sigma} \newc_{i,-\sigma})  \\
\label{H_TJ_ROT}
    &&+ \, {J \over 8} \sum_{\langle ij\rangle} \ 
       \left (
       \newS_i^- \newS_j^+ \,-\, 3 \newS_i^+ \newS_j^+  + {\rm h.c.}
       \right ) \nonumber \\
    &&+ \, {J \sqrt{3} \over 4} \sum_{\langle i\rightarrow j\rangle} \ 
       \left [
       \newS_i^z (\newS_j^+ + \newS_j^-) \,-\, \newS_j^z (\newS_i^+ + \newS_i^-)
       \right ] 
\,.
 \nonumber 
\end{eqnarray}
$\newc^{(\dagger)}$ are the electron destruction (creation) operators in
the rotated basis, $\newS$ are the corresponding spin operators. 
Like in the original $t$-$J$ model the operators $\newc^{(\dagger)}$
exclude double occupancies.
The summation symbol $\langle i\rightarrow j\rangle$ refers to orientated 
pairs $\langle ij \rangle$ where the link $i \rightarrow j$ runs in positive 
direction with respect 
to one of the unit vectors ${\bf e}_i$ of the lattice.
The ground state of $H_0$ is now ferromagnetic, whereas $H_1$ contains the 
fluctuations.
An important difference compared to the square-lattice hole motion problem 
is the existence of a direct hopping term (in $H_0$), i.e., hopping without 
creation of a background spin-defect. This follows
from the fact that the spin states on adjacent sites in the classical
ground state are not exactly orthogonal.

The hole motion processes will be described in the concept of path operators
~\cite{Strings,Trugman88,Vojta98}
which create strings of spin fluctuations attached to the hole.
%
For the application of projection technique we define a set of path operators
$\{A_I\}$ which couple to a hole and create local spin defects with respect to
the classical ground state.
The first operator $A_0$ is the unity operator, the second one $A_1$ moves the hole
by one lattice spacing creating one spin defect and so on.
We are interested in calculating dynamical correlation functions for the
operators $\{A_I \oldc_{{\bf k}\sigma}\}$:
\begin{equation}
G_{I\sigma,J\sigma'}(z)
       \,=\, \langle\psi_0^N|\,(A_I \oldc_{{\bf k}\sigma})^\dagger {1 \over {z-L}}
                               (A_J \oldc_{{\bf k}\sigma'})^{\phantom\dagger}\, 
               |\psi_0^N\rangle \,.
\label{ARBKORR}
\end{equation}
The Green's function (\ref{HOLE_GF}) for the physical hole is then 
given by $\sum_\sigma G_{0\sigma,0\sigma}(z)$. 
Note that it is equivalent to consider correlation functions for operators
$\{A_I \newc_{{\bf k}\sigma}\}$; then the physical Green's function (\ref{HOLE_GF}) has 
to be calculated using relation (\ref{OLDNEWTRAFO}).
Using cumulants the correlation functions $G$ can be rewritten as \cite{BeckBre90}:
\begin{equation}
G_{I\sigma,J\sigma'} (z) \,=\,
  \langle\phi_{\rm cl}|\,\Omega^\dagger\,
  (A_I \oldc_{{\bf k}\sigma})^\dagger \left( {1 \over {z-L}}
  A_J \oldc_{{\bf k}\sigma'}^{\phantom\dagger} \right)^{\cdot} \, \Omega\,
  |\phi_{\rm cl}\rangle^c
\,.
\label{KUMKORR}
\end{equation}
The brackets $\langle\phi_{\rm cl}|\,...\,|\phi_{\rm cl}\rangle^c$ denote cumulant expectation
values with $|\phi_{\rm cl}\rangle$.
The dot $\cdot$ in Eq. (\ref{KUMKORR}) indicates that the quantity 
inside $(...)^{\cdot}$ 
has to be treated as a single entity in the cumulant formation.
The operator $\Omega$ transforms $|\phi_{\rm cl}\rangle$ 
being the ground state of $H_0$ into the exact ground state $|\psi_0^N\rangle$ of 
$H = H_0+H_1$ at half-filling.
Here $\Omega$ is approximated with an exponential
ansatz which introduces spin fluctuations into $|\phi_{\rm cl} \rangle$:
\begin{eqnarray}
|\psi_0^N\rangle \:=\: \Omega\, |\phi_{\rm cl}\rangle \:=\:
 \exp \left(\sum_\nu \alpha_\nu S_\nu \right) 
 |\phi_{\rm cl}\rangle
\,.
\label{OMEGA_QAFM}
\end{eqnarray}
The operators $S_\nu$ describe the effect of the spin-flip terms in $H_1$.
This approach has been shown to give reasonable results for the square lattice 
AF \cite{BeckFul88}.
Here operators with up to 4 spin defects with a maximum
distance of 4 lattice spacings have been employed.
Following ref. \cite{SchorkFul92} one obtains a non-linear set of equations
for the coefficients $\alpha_\nu$, 
$0 = \langle\phi_{\rm cl}| S_\nu^\dagger\, H \, \Omega\, |\phi_{\rm cl}\rangle^c$,
which can be solved self-consistently.

Using Mori-Zwanzig projection technique \cite{Projtech} one can derive
a set of equations of motion for the dynamical correlation
functions $G_{I\sigma,J\sigma'}(z)$.
Neglecting the self-energy terms it reads:
\begin{eqnarray}
\sum_{I\sigma}  \Omega_{K\tilde\sigma,I\sigma} (z)
  \,G_{I\sigma,J\sigma'}(z)\,\,=\,\,\chi_{K\tilde\sigma,J\sigma'} \, ,
  \nonumber \\
\Omega_{K\tilde\sigma,J\sigma'}(z) =
  z \delta_{KJ}\delta_{\tilde\sigma\sigma'}
  - \sum_{L\sigma''}\omega_{K\tilde\sigma,L\sigma''}^{\phantom{-1}}\,\chi_{L\sigma'',J\sigma'}^{-1}
\,.
\label{PROJ_GLSYS}
\end{eqnarray}
$\chi_{I\sigma,J\sigma'}$ and $\omega_{I\sigma,J\sigma'}$
are the static correlation functions and frequency terms,
respectively.
They are given by the following cumulant expressions:
\begin{eqnarray}
\chi_{I\sigma,J\sigma'} &=& 
  \langle\phi_{\rm cl}| \Omega^\dagger
  (A_I \oldc_{{\bf k}\sigma})^{\cdot \dagger}
  (A_J \oldc_{{\bf k}\sigma'})^\cdot \Omega \,
  |\phi_{\rm cl}\rangle^c \, , \nonumber\\
\omega_{I\sigma,J\sigma'} &=& 
  \langle\phi_{\rm cl}| \Omega^\dagger
  (A_I \oldc_{{\bf k}\sigma})^{\cdot \dagger}
  \left(L (A_J \oldc_{{\bf k}\sigma'})^{\phantom\dagger} \right)^\cdot \Omega \,
  |\phi_{\rm cl}\rangle^c
\,.
\label{QAFM_MATEL}
\end{eqnarray}
These terms describe all dynamic processes within the subspace of the
Liouville space spanned by the operators $\{A_I\oldc_{{\bf k}\sigma}\}$.
The use of cumulants ensures size-consistency, i.e., only spin fluctuations connected
with the hole enter the final expressions for the one-hole correlation 
function.
In the present calculations we have employed up to 900 projection variables with 
a maximum paths length of 4.
The neglection of the self-energy terms leads to a discrete set of poles for
the Green's functions, in all figures we have introduced
an artificial linewidth to plot the spectra.
For details of the calculational procedure see e.g. \cite{Vojta98}.



\begin{figure}
\epsfxsize=8.2 truecm
\centerline{\epsffile{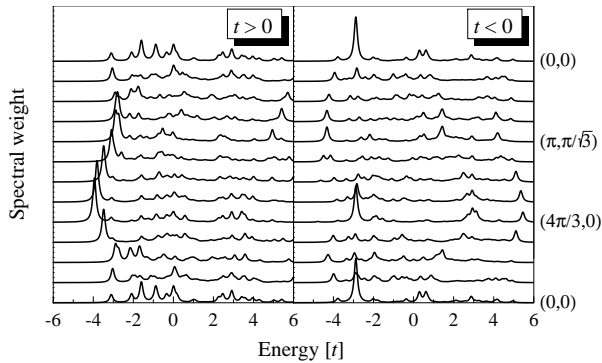}}
\caption{
  One-hole spectral function for $J/|t| = 0.5$ and different momenta;
  left: positive $t$, right: negative $t$.
  The energies are measured relative to the energy of a localized hole;
  the path in the Brillouin zone is shown in Fig. 1.
}
\label{FIG_SPEC1}
\end{figure}

\begin{figure}
\epsfxsize=8.2 truecm
\centerline{\epsffile{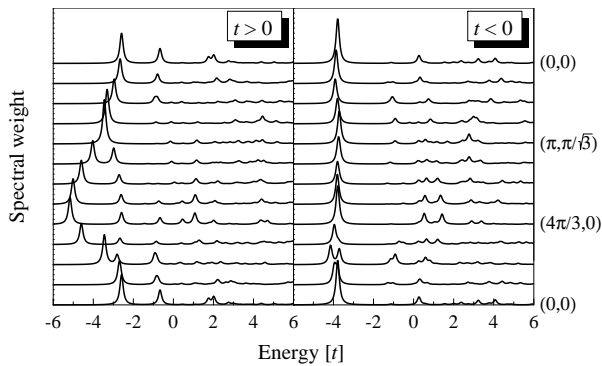}}
\caption{Same as Fig. 2 but for $J/|t| = 2.0$.}
\label{FIG_SPEC2}
\end{figure}

\begin{figure}
\epsfxsize=8.2 truecm
\centerline{\epsffile{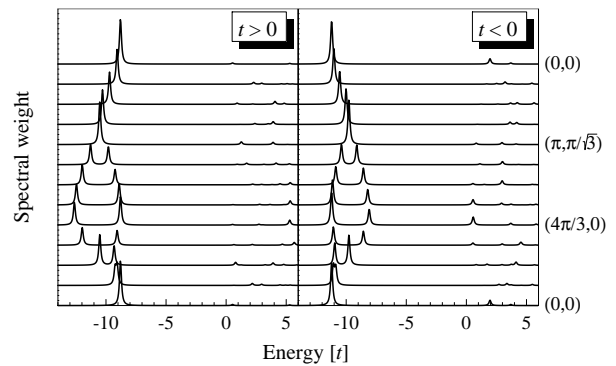}}
\caption{Same as Fig. 2 but for $J/|t| = 10.0$.}
\label{FIG_SPEC3}
\end{figure}

Now we turn to the discussion of the results.
The one-hole spectral function ${\rm Im}\,G({\bf k},\omega)$ 
for both positive and negative $t$ and different momenta is shown in 
Figs. \ref{FIG_SPEC1} -  \ref{FIG_SPEC3} ($J/t =$ 0.5, 2.0, and 10.0).
For large $J/t$ we observe pronounced low-lying QP bands and additional excitations with 
low weight at higher energies which form an incoherent background. 
With decreasing $J/t$ the gap between the QP band and the background excitations
decreases. 
For $J/t=0.5$ this gap is almost vanished, furthermore a sharp QP peak is only
present near the bottom of the QP band.
Values of $J/t \leq 0.2$ lead to a incoherent spectrum.

For negative $t$ the character of the spectral function is in principle similar
to the $t>0$ results, i.e., for large $J/|t|$ ($\geq 1.0$) the QP peak is well defined, whereas for
small $J/|t|$ the spectra become incoherent.

\begin{figure}
\epsfxsize=8.2 truecm
\centerline{\epsffile{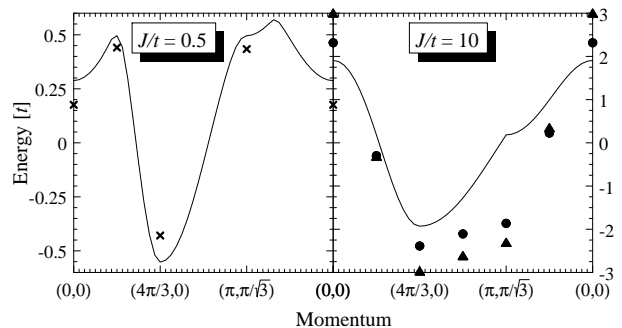}}
\caption{ 
  Dispersion of the low-lying QP band for $t>0$.
  Left: $J/|t| = 0.5$, present calculation (solid) and ED data (12 sites, crosses) 
  taken from ref. \cite{Azzouz}. 
  Right: $J/|t| = 10$, present calculation (solid) together SCBA data (dots)
  and ED data (21 sites, triangles) from ref. \cite{Apel}. 
  The energy zero level has been set at the center of mass of the band. 
}
\label{FIG_ONEHOLE}
\end{figure}

Next we are going to examine the properties of the low-lying bands.
Fig. \ref{FIG_ONEHOLE} shows the dispersion $\epsilon_{\bf k}$ of the lowest pole
for positive $t$ and different $J/t$ in comparison with data
from the literature. 
For $t>0$ the band has its energy minimum at momentum $({4 \over 3}\pi,0)$
and equivalent points. 
For $t<0$ and large $J/|t| \geq 4$ the situation is similar to the positive $t$
case with the dispersion of the QP band reversed, see Fig. \ref{FIG_SPEC3}.
However, for intermediate $J/|t|$ and $t<0$ the QP dispersion almost vanishes
(Fig. \ref{FIG_SPEC2}). 
This can be explained from the composite nature of the hole motion 
process in the triangular system. 
The motion consists of direct hopping [amplitude $t/2$, cf. eq. (\ref{H_TJ_ROT})]
and spin-fluctuation-assisted hopping (like in the square lattice)
with an amplitude being nearly independent of
the sign of $t$ (since the main contribution contains two hopping steps and one 
spin-flip process).
For negative $t$ these two contributions to the dispersion tend to cancel each other
leading to the very narrow band at $J/|t| = 2.0$.

Another interesting feature is the splitting of the QP band into two which is
especially visible at large $J/|t|$.
It is related to the non-collinear magnetic long-range order in the system
and can be understood as follows:
The states $|\phi_{\rm cl}\rangle$ and $|\phi_0^N\rangle$ have a non-zero 
helicity, therefore both $y$ directions in spin space (perpendicular to 
the plane defined by the spin directions in $|\phi_{\rm cl}\rangle$)
are not equivalent.
The analysis of the eigenvectors of the dynamic matrix $\Omega_{I\sigma,J\sigma'}$
shows that the one-hole eigenmodes correspond to states 
where the missing spin has a definite $y$ component.
(These are no eigenstates of $S_z^{\rm tot}$.)
These two modes have different energies (for a given ${\bf k}$) which leads to two distinct bands 
in the spin-integrated spectral function (\ref{HOLE_GF}). 
A spin-resolved photoemission experiment (with spins parallel or antiparallel to the 
helicity direction) should observe one {\bf or} the other of these bands.


The one-hole dispersion for $J/t=0.5$ calculated here is in 
quantitative agreement with exact diagonalization (ED) data on a 
12-site cluster~\cite{Azzouz}.
The $J/t=10$ dispersion curve agrees with recent results which where obtained
using the SCBA technique and ED on a 21-site cluster~\cite{Apel} 
(see Fig. \ref{FIG_ONEHOLE}). 
To compare results, one has to take into account an overall shift of the hole momenta 
by $({4 \over 3}\pi,0)$ between clusters with odd and even number of sites.

\begin{figure}
\epsfxsize=7 truecm
\centerline{\epsffile{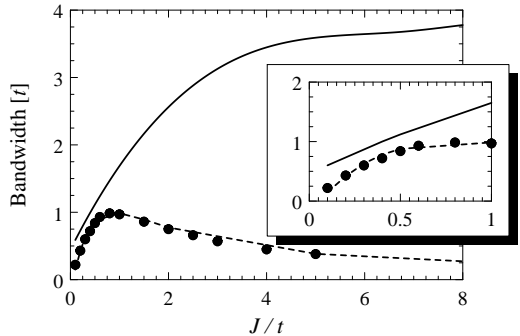}}
\caption{Comparison of the hole QP bandwidth in triangular and square lattice
antiferromagnets. 
Solid/dashed: Triangular/square lattice, present calculation. 
Circles: Square lattice SCBA results from ref. \cite{MaHo}.
Inset: $J$ range from 0 to 1.
}
\label{FIG_BWCOMP}
\end{figure}

Finally we want to compare the QP bandwidth with 
the one found for the square lattice hole motion.
In Fig. \ref{FIG_BWCOMP} we have plotted the bandwidth for both systems
depending on the ratio $J/t$.
For the square lattice it is known \cite{Riera97,Vojta98,MaHo}
that the bandwidth for small $J/t$ ($< 1$) is essentially given
by $J$ since hole hopping always creates spin defects with respect to the
AF background. These defects have to be "repaired"
by the transverse part of the Heisenberg exchange with energy scale $J$.
In contrast, in the triangular lattice the bandwidth for small $J/t$ 
is given by $a t + b J$, whereas for large $J/t$ values it saturates at 
$4.5\,t$.
This behavior follows from the existence of the direct hopping term 
with prefactor $t/2$ in $H_0$ (\ref{H_TJ_ROT}) which has been discussed 
above.
It leads to a disperison proportional to $t$; the saturation bandwidth
is half of the bandwidth of the uncorrelated system ($9\,t$).
Additional hole motion processes accompanied by
spin fluctuations lead to a dispersion proportional to $J$ (at small $J/t$) as 
in the square lattice. 
So the physical picture for the hole quasiparticle is the following:
At small and intermediate $J/t$ the hole is surrounded by spin fluctuations,
the dynamics consists of coherent quasiparticle motion and of incoherent 
processes within the quasiparticle (which dominate the spectrum at small 
$J/t$).
At large $J/t$ spin fluctuations are suppressed, but the direct hopping term
allows the hole to move without creating background spin defects. 
So the pure hole hops with $t/2$ as a (nearly) free fermion.



Summarizing, we have studied the one-hole motion in a
triangular AF described by the $t$-$J$ model.
Using the spin-polaron concept which describes the hole states in
terms of local spin defects we find that the one-hole spectral
function shows a QP peak for sufficiently large $J/|t|$.
In this regime the picture of a mobile hole dressed by spin fluctuations 
(known from the square lattice) also applies to the triangular 
system. 
The QP dispersion arises both from direct and spin-flip-assisted hopping 
processes which leads to differences between the $t>0$ and $t<0$
cases.
For small $J/|t|$ the one-hole spectra become incoherent, i.e., the QP 
weight decreases and the gap between the QP band and the background
formed by additional magnetic excitations vanishes.

Note that the addition of more than one hole to a triangular AF
may lead to hole pairing induced by spin-wave exchange~\cite{TriSC}
which is likely relevant for organic superconductors \cite{BEDTTTF}.

The author thanks K. W. Becker and E. Dagotto for useful conversations.
Support by the DAAD (D/96/34050) and the hospitality of the NHMFL
(Tallahassee) are gratefully acknowledged. 



\end{multicols}


\begin{references}

\bibitem{Dagotto94}E. Dagotto, Rev. Mod. Phys. {\bf 66}, 763 (1994).
\bibitem{Strings}Y. Nagaoka, Phys. Rev. {\bf 147}, 392 (1966), 
        W. F. Brinkman and T. M. Rice, Phys. Rev. B {\bf 2}, 1324 (1970), 
        B. I. Shraiman and E. D. Siggia, Phys. Rev. Lett. {\bf 60}, 740 (1988).
\bibitem{Trugman88}S. A. Trugman, Phys. Rev. B {\bf 37}, 1597 (1988).

\bibitem{Riera97}J. A. Riera and E. Dagotto, Phys. Rev. B {\bf 55}, 14543 (1997).
\bibitem{Vojta98}M. Vojta and K. W. Becker, Phys. Rev. B {\bf 57}, 3099 (1998).
\bibitem{MaHo}G. Martinez and P. Horsch, Phys. Rev. B {\bf 44}, 317 (1991).

\bibitem{Takane94} K. Hirakawa $et$ $al.$, J. Phys. Soc. Jpn. {\bf 54}, 3526
        (1985); K. Takeda $et$ $al.$, $ibid$ {\bf 61}, 2156 (1992).
\bibitem{Surfaces}
        H. H. Weitering $et$ $al.$, Phys. Rev. Lett. {\bf 78}, 1331 (1997).
\bibitem{defa} R. Cava $et$ $al.$, J. Solid State Chem. {\bf 104}, 437 (1993);
         A. Ramirez $et$ $al.$, Phys. Rev. B {\bf 49}, 16082 (1994).
\bibitem{BEDTTTF}
        R. H. McKenzie, 
        Science {\bf 278}, 820 (1997).

\bibitem{Fazekas74}P. Fazekas and P. W. Anderson, Phil. Mag. {\bf 30},
        423 (1974).
\bibitem{Singh92}R. R. P. Singh and D. Huse, Phys. Rev. Lett. {\bf 68},
        1766 (1992).
\bibitem{Bernu94}B. Bernu et $al.$, Phys. Rev. B {\bf 50}, 10048 (1994).

\bibitem{Azzouz}M. Azzouz and T. Dombre, Phys. Rev. B {\bf 53}, 402 (1996).
\bibitem{Apel}
        W. Apel, H.-U. Everts, and U. K\"{o}rner, Eur. Phys. J. B {\bf 5}, 317 (1998).

\bibitem{BeckFul88}K. W. Becker and P. Fulde, Z. Phys. B {\bf 72}, 423 (1988).
\bibitem{BeckBre90}K. W. Becker and W. Brenig, Z. Phys. B {\bf 79}, 195 (1990).

\bibitem{Projtech}H. Mori, Progr. Theor. Phys. {\bf 34}, 423 (1965), 
        R. Zwanzig, in: Lectures in
        Theoretical Physics \ vol. 3. New York: Interscience 1961.

\bibitem{SchorkFul92}T. Schork and P. Fulde, J. Chem. Phys. {\bf 97}, 9195 (1992).

\bibitem{TriSC}
        M. Vojta and E. Dagotto, preprint cond-mat/9807168,
        J. Schmalian, Phys. Rev. Lett. {\bf 81}, 4232 (1998).
         
\end{references}
\end{document}